\documentclass[a4paper]{VisionStyle}
\usepackage{epsfig}

\begin{document}

\title{XMM-Newton observations of the Lockman Hole : Spectral analysis}

\author{V.\,Mainieri\inst{1,2} \and J.\,Bergeron\inst{3} \and P.\,Rosati\inst{1} 
\and  G.\,Hasinger\inst{4,5} \and I.\,Lehmann\inst{4,5} } 

\institute{
  European Southern Observatory, Karl-Schwarzschild-Strasse 2, 
  D-85748 Garching, Germany
\and 
  Dipartimento di Fisica, Universit\`a degli Studi Roma Tre,
  Via della Vasca Navale 84, I-00146 Roma, Italy
\and 
  Institut d'Astrophysique de Paris, CNRS, 98 bis Boulevard Arago, F-75014 Paris, 
  France 
\and
 Max-Planck-Institut f\"ur Extraterrestrische Physik, Giessenbach-Strasse 
 Postfach 1312, D-85741  Garching, Germany  
\and
 Astrophysikalisches Institut Potsdam, An der Sternwarte 16, D-14482 Potsdam, 
 Germany  
}

\maketitle 

\begin{abstract}

We present the results of the X-ray spectral analysis of the deep 
survey obtained with the XMM-Newton observatory on the Lockman Hole.  
The X-ray data and the cumulative source counts were reported by 
\cite*{JB-E1:has01}. Our sample contains 104  sources with a 
count limit of 70  of which 55  have redshift identification. 
The redshift distribution peaks at $z \sim$ 0.8, with a strong excess of 
low $z$ AGN and a deficiency of sources at $z >$ 2 compared to population 
synthesis models for the X-ray background. The type 2 (obscured)  AGN 
have weaker soft X-ray and optical fluxes. They cluster around $z \sim$ 1. 
There is a clear separation between the classical/type 1 AGN and the 
obscured/type 2 ones in several diagnostics involving X-ray colour, X-ray  
flux, optical/near IR colour and optical brightness. Using the $z$ subsample, 
we show that this separation between the AGN populations is a consequence 
of different absorption column densities. The two populations have the same 
average spectral index,   $\langle\Gamma\rangle \sim 1.9$. 
At the 70 count detection limit,  
there is also a strong overlap between the two populations in hard X-ray flux 
and near IR  brightness. These diagnostics should enable the classification 
of obscured/type 2 AGN very faint optically.  
\end{abstract}

\section{Introduction}
  
Recent deep X-ray surveys with the XMM-Newton and Chandra observatories have
revealed an important population of obscured (type 2) active galactic nuclei
(AGN) (\cite{JB-E1:has01}, \cite{JB-E1:bar01}, 
\cite{JB-E1:hor01}, \cite{JB-E1:toz01}, \cite{JB-E1:ros01}). 
Objects of this class were already detected in the very deep ROSAT survey 
of the  Lockman Hole region, with a few heavily obscured/type 2 AGN in the 
redshift range 1-3 and several type 2 AGN at $z < 1$ (\cite{JB-E1:leh01}). 
This population accounts for the progressive hardening of the average X-ray 
spectrum towards fainter fluxes (see e.g. \cite{JB-E1:toz01}).
From their source number counts in the hard X-ray band, \cite*{JB-E1:gia01}
 and \cite*{JB-E1:ros01} showed that a very significant fraction 
($\sim$ 85-90\%) of the 2-10 keV X-ray background is resolved. 

The first deep X-ray survey with XMM-Newton was obtained during Performance
Verification. The observed field was centered on the  Lockman Hole (LH),   
RA 10:52:43 and DEC +57:28:48 (J2000), and the exposure time for the good 
quality observations was $\simeq$ 100 ksec. The X-ray data reduction and
analysis (restricted to sources within a 10$\arcmin$ radius)
was reported by \cite*{JB-E1:has01} who show, in particular, that the 
sources could be classified by their X-ray colours. 

The aim of this paper is to perform an X-ray  spectral analysis of the 
LH sources and characterize their X-ray and optical/near IR properties. 
We thus search for relations between X-ray and/or optical physical 
parameters in the full  X-ray sample and, using the subsample with redshift 
identification, check the validity of our conclusions concerning the 
specific properties of the obscured/type 2 AGN population.

\section{The X-ray sample}
\label{JB-E1_sec:sam}

From the deep XMM-Newton pointing, we selected the sample of 104 X-ray 
sources with a  number of counts larger than 70 in the 0.5-10 keV energy 
band, of which 70 sources are within an off-axis angle of 10$\arcmin$.  
This minimum number 
of counts is chosen as a compromise between the sample size and the 
accuracy of the X-ray spectral fit. Of these 104 sources, 
60 were not detected by the ROSAT observatory. The SAS source detection 
algorithm was applied to PN data  and the count extraction 
was performed using the Sextractor algorithm. The latter allows a
determination of the ellipsoidal shape of the source (important tool, in 
particular for the outer regions of the field where there are strong 
distorsions of the PSF) and an easy subtraction of the 
neighbouring sources for the estimate of the source net counts. A detailed 
presentation of the data reduction is given in \cite*{JB-E1:mai02}. 

The Lockman Hole field was already extensively studied with ROSAT 
(\cite{JB-E1:has98}) and ASCA (\cite{JB-E1:ish01}) as well as at other 
wavelengths, including optical spectroscopic identification 
(see \cite{JB-E1:sch98}, \cite{JB-E1:leh01} and references therein). 
Excluding of the sample the two galaxy clusters and the eight stars, there 
remains 94 X-ray sources of which 55 have redshift identification.

\section{The redshift distribution}
\label{JB-E1_sec:red}

Synthesis models for the X-ray background, using observational constraints 
mostly from ROSAT and ASCA surveys, predict a substantial population of 
type 2 AGN and a fairly broad redshift distribution which peaks at 
$z \sim 1.5$ (\cite{JB-E1:gil01}). 
In order to compare our results with those predicted by these authors, we 
use an extended sample of 119 AGN with a count limit of 50, of which 73 have 
known redshifts (completeness level of 61\%). This count limit corresponds 
to a flux in the 0.5-2 keV band which is a factor of 2 higher than the models 
flux limit of $2.3 \times 10^{-16}$ erg cm$^{-2}$ s$^{-1}$.   
From the log N-log S distribution given by \cite*{JB-E1:toz01}, 
we then estimate  that there is a factor of $\sim 1.8$ in the cumulative 
counts between these two flux limits. The conclusions drawn below should 
thus be viewed as preliminary.

\begin{figure}[!ht]
  \begin{center}
    \psfig{file=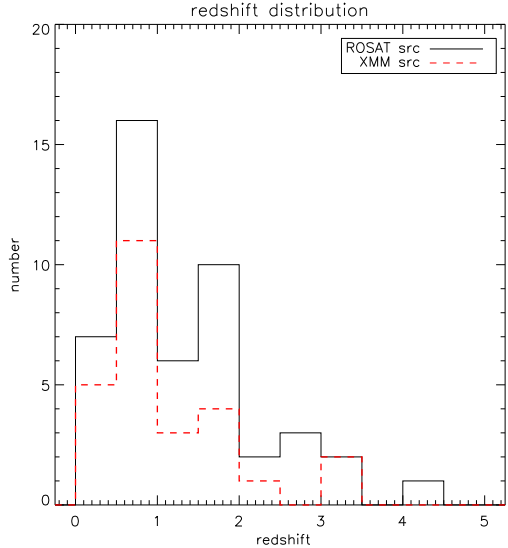, width=7.5cm}
  \end{center}
\caption{Redshift distribution of the X-ray sources in the Lockman Hole. 
Separate histograms are given for the sources detected by XMM-Newton only 
and those also detected by ROSAT.}  
\label{JB-E1_fig:fig1}
\end{figure}

The $z$ distribution of the LH sources is shown in 
Figure~\ref{JB-E1_fig:fig1}. In our $z$ subsample, there are 53\% of the 
sources at $z < 1$ and only 15\% at $z > 2$. This distribution, which peaks at 
$z \sim 0.8$, markedly differs from the predicted ones, with a strong excess 
of low $z$ AGN and a deficiency of high $z$ ones. Even if all the LH 
unidentified sources were at $z >$ 1, there would still be 33\% of the LH 
sources at  $z <$ 1 compared to the 10.5-15\% in the models. 
Such a discrepancy between predictions and observations was already noted 
by \cite*{JB-E1:ros01} for the 1 Msec Chandra survey of the Chandra 
Deep Field South (CDFS).

\section{Spectral analysis}
\label{JB-E1_sec:xsp}

We use xftools to extract individual spectra and perform the spectral 
analysis. In this preliminary study, we assume a simple model for the
spectral fit: a single power law of photon index $\Gamma$ and a column 
density $N_{\rm H}$. In the analysis presented by \cite*{JB-E1:mai02},
an additional component is introduced when there is an unambiguous soft 
X-ray excess.  The minimum value of $N_{\rm H}$ is that of the low 
Galactic Hydrogen column density towards the LH field, 
$N_{\rm H}=5.7 \times 10^{19}$ cm$^{-2}$ (\cite{JB-E1:loc86}).

We use the results of the spectral analysis, in particular the subsample 
with $log N_{\rm H}> 21.5$, to confirm the different properties of the 
classical and obscured AGN populations when studying the whole X-ray 
sample.

\subsection{Hardness ratios}
\label{JB-E1_sec:hr}

X-ray colours are usually represented by hardness ratios, $HR=(H-S)/(H+S)$ 
where $H$ and $S$ correspond to the counts in the harder and softer energy 
bands respectively. 
In the X-ray colour-colour diagrams presented by \cite*{JB-E1:has01},
there is a clear separation of the type 1 and 
type 2 ROSAT sources. Moreover, a large fraction of the XMM-Newton only 
sources, typically fainter than the ROSAT ones, fill the same colour-colour 
space than the ROSAT type 2 sources. 

\begin{figure}[!hb]
  \begin{center}
    \psfig{file=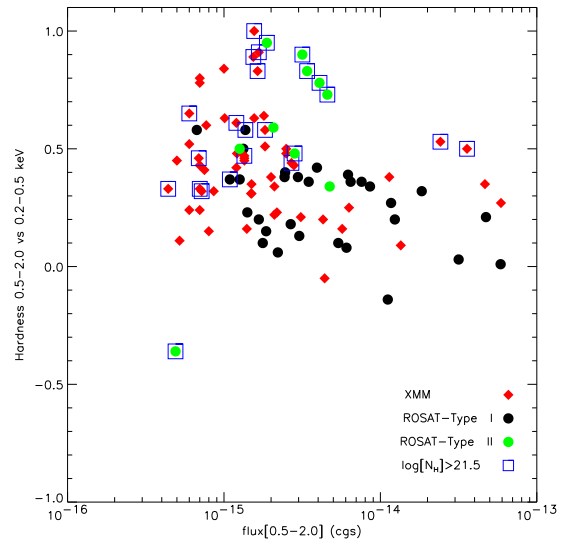, width=7.5cm}
  \end{center}
\caption{Hardness ratio (0.5-2 keV vs 0.2-0.5 keV) 
versus X-ray flux in the band 0.5-2 keV.
}  
\label{JB-E1_fig:fig2}
\end{figure}

\begin{figure}[!ht]
  \begin{center}
    \psfig{file=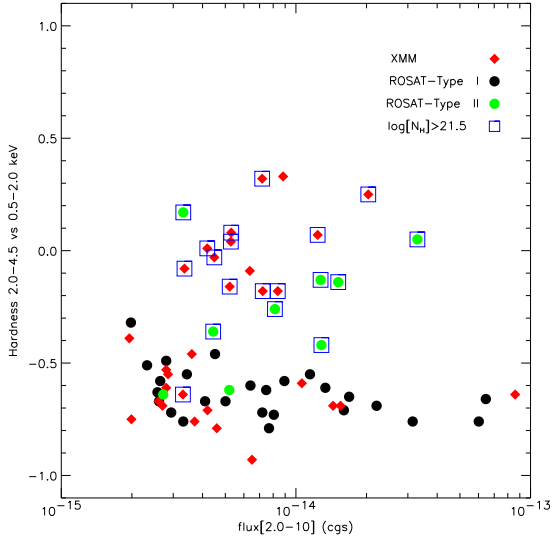, width=7.5cm}
  \end{center}
\caption{Hardness ratio (2-4.5 keV vs 0.5-2 keV) versus  
X-ray flux in the band 2-10 keV.
}  
\label{JB-E1_fig:fig3}
\end{figure}

We consider X-ray colour versus X-ray flux diagrams  to identify which 
ones show a clearer separation between the different AGN populations.
The latter  are presented in Figures~\ref{JB-E1_fig:fig2} and  
\ref{JB-E1_fig:fig3}. The obscured sources  with $log N_{\rm H}> 21.5$ 
have specific labels. For the type 1 AGN, we get 
$\langle HR$(0.5-2 keV vs 0.2-0.5 keV)$\rangle \sim 0.25$  
 whereas for most of the type 2 AGN we get values of this $HR$ larger than 
0.5. The large majority of the obscured sources overlap with the type 2 AGN. 
One should note that there are two type 1 AGN with large 
$N_{\rm H}$ and some of the obscured XMM-Newton sources with similar 
values of $HR$  could also belong to that subclass of objects. 

An even clearer diagnostic is obtained with  $HR$(2-4.5 keV vs 0.5-2 keV) 
shown as a function of $F$(2-10 keV). The type 1 AGN cluster around 
$\langle HR$(2-4.5 keV vs 0.5-2 keV)$\rangle \sim -0.6$, but the type 2 AGN, 
as well as all the obscured sources but one, span a large $HR$ range 
from $-0.4$ to +0.3.

These results demonstrate that there is a marked separation between the 
two AGN populations in the  X-ray colour versus X-ray flux diagrams, and 
 that this is mostly due to differences in the absorption column 
density.

\subsection{Spectral index and column density}
\label{JB-E1_sec:gnh}

\begin{figure}[!ht]
  \begin{center}
    \psfig{file=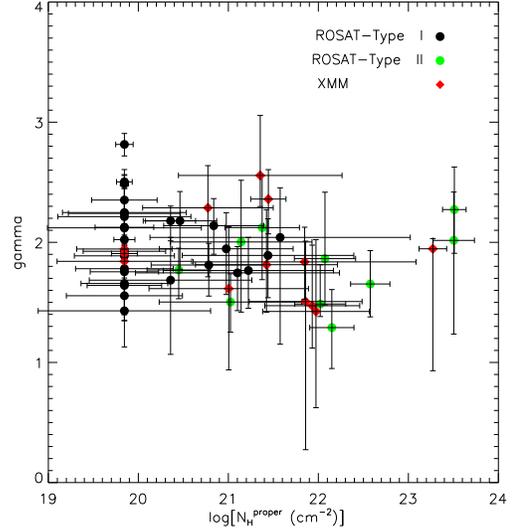, width=7.5cm}
  \end{center}
\caption{Rest-frame photon spectral index versus absorption column density 
for the $z$ subsample of LH sources.
}  
\label{JB-E1_fig:fig4}
\end{figure}

\begin{figure}[!ht]
  \begin{center}
    \psfig{file=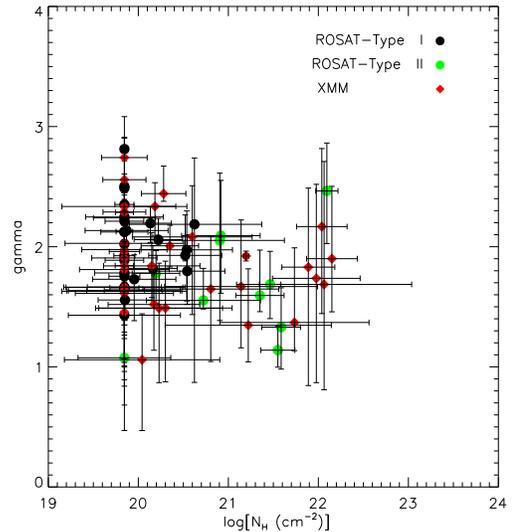, width=7.5cm}
  \end{center}
\caption{Observed photon spectral index versus absorption column density for 
the whole LH sample.
}  
\label{JB-E1_fig:fig5}
\end{figure}

The spectral analysis is performed for both the $z$ subsample, which leads 
to proper values of $\Gamma$ and $N_{\rm H}$ in the source rest-frame, and 
the complete sample for which observed values of $\Gamma$ and $N_{\rm H}$ 
are then obtained. Results are shown in Figures~\ref{JB-E1_fig:fig4} 
and \ref{JB-E1_fig:fig5}.

In both diagrams, the type 1 and 2 AGN populations occupy different loci in 
$N_{\rm H}$. For the $z$ subsample, the bimodal distribution appears to also 
apply to $\Gamma$. This result is most probably an artefact due to the
presence of a strong soft X-ray excess in several of the type 2 AGN. When 
this additional component is introduced in the spectral analysis, the values 
obtained for both $\Gamma$ and $N_{\rm H}$ are then larger (see 
\cite{JB-E1:mai02}). 

The average values of $\Gamma$ derived for both the $z$ subsample and the 
complete sample are similar, $\langle\Gamma\rangle \sim 1.9$ with a large 
spread of $\pm 0.9$. The difference between the two samples is thus mainly 
in $N_{\rm H}$, 
not in $\Gamma$. These results differ from those obtained from observed, 
stacked spectra for which the signature of absorption column density is 
washed out (mainly a redshift effect) and the average value of $\Gamma$ 
is thus lower (\cite{JB-E1:toz01}). Our analysis confirms that the trend
found by these authors of a hardening of the stacked spectra (decrease of 
$\langle\Gamma\rangle$) with decreasing X-ray flux in the band 2-10 keV is
indeed a consequence of an increasing fraction of obscured sources at 
fainter flux levels.

From the optical identifications, we note that among the obscured sources  
of the $z$ subsample there are 3 type 1 AGN (of which one not detected by
ROSAT) and several XMM-Newton only sources. In the whole sample, the 
XMM-Newton only sources belong to several populations: type 2 AGN, EROs 
(R$-$K$ > 5$), and those with low values of $N_{\rm H}$ could be 
faint type 1 AGN or high $z$ type 2 AGN.

\section{X-ray and optical properties}
\label{JB-E1_sec:xop}

Another parameter, the optical/near IR colour, helps to distinguish between 
the two types of AGN as shown by \cite*{JB-E1:has99} and \cite*{JB-E1:leh01} 
in their analyses of the ROSAT deep survey of the Lockman Hole, and 
also recently by \cite*{JB-E1:bar01} and \cite*{JB-E1:ros01} in their 
studies of the Chandra deep surveys of the  CDFN and CDFS respectively.
We perform a similar analysis on the XMM-Newton LH field data.

\subsection{Optical colours}
\label{JB-E1_sec:col}

\begin{figure}[!hb]
  \begin{center}
    \psfig{file=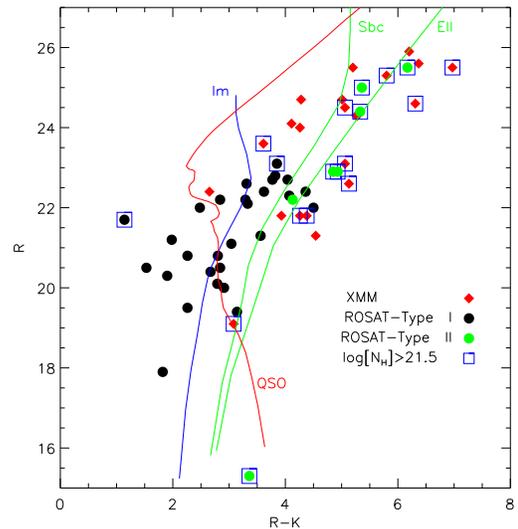, width=7.5cm}
  \end{center}
\caption{Colour-magnitude diagram, R versus R$-$K, for sources of the LH 
field. See text for explaination of the evolutionary tracks.
}  
\label{JB-E1_fig:fig6}
\end{figure}

\begin{figure}[!hb]
  \begin{center}
    \psfig{file= 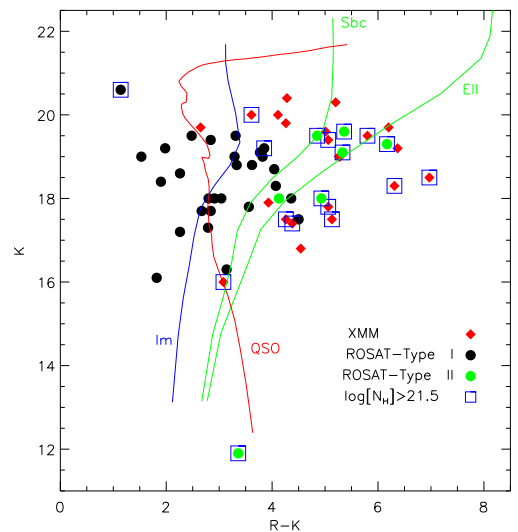, width=7.5cm}
  \end{center}
\caption{Colour-magnitude diagram, K versus R$-$K, for sources of the LH field.
}  
\label{JB-E1_fig:fig7}
\end{figure}

In Figures~\ref{JB-E1_fig:fig6} and \ref{JB-E1_fig:fig7}, we present
the R$-$K colour versus R and K respectively.
Data are available  for about one half of the 70
count limit sample. The evolutionary tracks shown for the various types of 
galaxies are taken from \cite*{JB-E1:col80}. Their spectral energy 
distributions (SED) were extended to the near-IR using the models of 
\cite*{JB-E1:bc93} as updated in 2000 (private communication). The QSO 
evolutionary track is derived from the empirical template from the Sloan 
Digital Sky Survey (\cite{JB-E1:van01}), together with the models of 
\cite*{JB-E1:gra97}, normalized to $M_B^{\ast}=-22.4$, for the 
extension in the near IR. 

The classical AGN roughly cluster around the QSO evolutionary track, whereas
the type 2 and obscured AGN are clearly fainter and 
redder objects, most with R $\ga$ 22 and R$-$K $>$ 4.
The nuclei of the latter, being obscured, do not substantially contribute to 
the total R magnitude, nucleus + host. These objects cluster 
around the E and Sbc galaxy evolutionary tracks and are thus embedded 
in galaxies of various morphological types.  These conclusions are in full
agreement with the results of HST imaging of the CDFS for their ``brigther'' 
population (\cite{JB-E1:sch01}, \cite{JB-E1:koe01}). 

\begin{figure}[!hb]
  \begin{center}
    \psfig{file=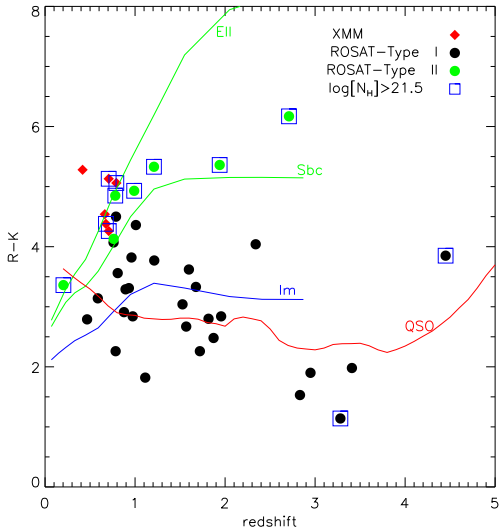, width=7.5cm}
  \end{center}
\caption{Colour R$-$K versus redshift for sources of the LH field.
}  
\label{JB-E1_fig:fig8}
\end{figure}

Although present, the separation between the different AGN populations in 
the R$-$K versus K colour diagram is not as striking as in the previous one. 
There is a strong overlap in K magnitude between the two populations.  
Although the type 2 and obscured AGN populations still roughly cluster 
around E and Sbc SED evolutionary tracks, the scatter is larger. 

The relation between colour, R$-$K, and redshift is shown in  
Figure~\ref{JB-E1_fig:fig8}. The type 1 AGN span a wide redshift range 
whereas the type 2 AGN, together with the majority of the obscured ones, 
cluster around $z \sim$ 1 as already noted for the CDFS sources 
(\cite{JB-E1:ros01}).

\subsection{X-ray and optical fluxes}
\label{JB-E1_sec:flu}

To further characterize the properties of the different X-ray populations, 
we investigate the relations between their X-ray flux and optical magnitude.

\begin{figure}[!ht]
  \begin{center}
    \psfig{file=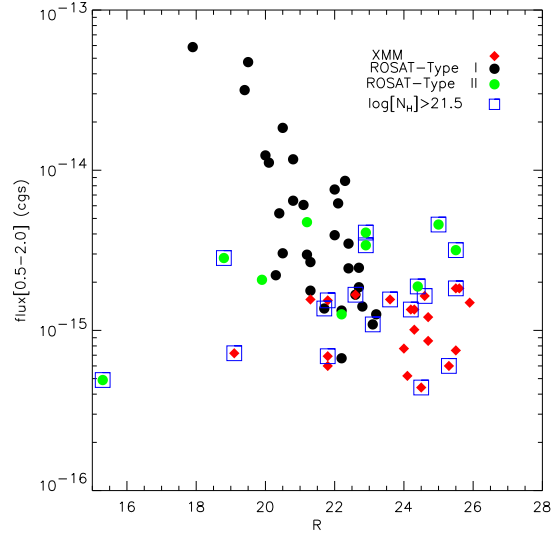, width=7.5cm}
  \end{center}
\caption{Soft X-ray flux versus R magnitude for the LH sources with optical
detections.
}  
\label{JB-E1_fig:fig9}
\end{figure}

\begin{figure}[!ht]
  \begin{center}
    \psfig{file=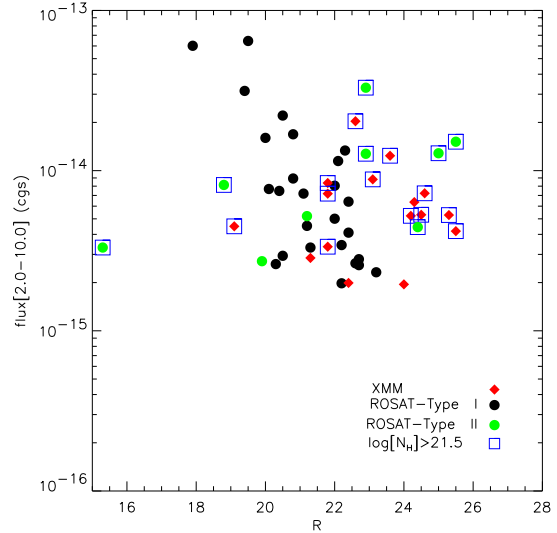, width=7.5cm}
  \end{center}
\caption{Hard X-ray flux versus R magnitude for the LH sources with optical
detections.
}  
\label{JB-E1_fig:fig10}
\end{figure}

As could be seen in Figure~\ref{JB-E1_fig:fig9}, there is a clear 
trend of increasing soft X-ray flux with optical brightness, this correlation
being tighter for the type 1 AGN. For classical quasars, the correlation 
between X-ray and optical emission is indeed known since the first quasar 
surveys with the Einstein observatory (\cite{JB-E1:zam81}).
The type 2 and obscured AGN, as well as 
the XMM-Newton only sources, have fainter soft X-ray fluxes and R magnitudes 
than the type 1 AGN and a fairly large scatter in both parameters. 

This separation in X-ray flux is no longer present for the hard X-ray band 
(see Figure~\ref{JB-E1_fig:fig10}). The trend observed in the previous
diagram is indeed a consequence of higher absorption column densities for 
type 2 AGN, as well as most XMM-Newton only sources,  and not smaller 
broad-band X-ray fluxes.

\section{Conclusions}
\label{JB-E1_sec:con}

We have derived the X-ray spectral properties of the sources detected by 
XMM-Newton in the Lockman Hole. The X-ray data reduction, analysis of the 
source counts and X-ray colour-colour diagnostics were reported by 
\cite*{JB-E1:has01}. The sample contains 104 sources at a count 
limit of 70, which corresponds to a flux of 
(0.69 and 3.9) $\times 10^{-15}$ erg cm$^{-2}$ s$^{-1}$ in the 0.5-2 and 
2-10 keV band respectively. It comprises 94 AGN of which 55 have 
redshift identification. Making use of the $z$ subsample in several X-ray 
and/or optical diagnostics, we could characterize the different populations 
of AGN. Our main results are as follows.  
\begin{itemize}
  \item The clearer separation between the classical/type 1 AGN and 
the obscured/type 2 AGN is present in the diagrams X-ray colour, 
$HR$(2-4.5 keV vs 0.5-2 keV), versus hard (2-10 keV) X-ray flux, optical R 
magnitude versus optical/near IR colour, R$-$K, and soft (0.5-2 keV) X-ray 
flux versus R magnitude.
  \item In these diagnostic diagrams, most of the XMM-Newton only sources 
overlap with the ROSAT  obscured/type 2 AGN and the scatter for these two 
classes of objects is much larger than that for the classical/type 1 AGN. 
  \item The differences in parameter space between the type 1 and 
obscured/type 2 AGN is essentially due to the absorption column density.
  \item The average photon spectral index is  $\langle\Gamma\rangle \sim 1.9$ 
for both type 1 and 2 AGN. 
  \item The obscured/type 2 AGN population is fainter than the 
classical/type 1 AGN population in the soft X-ray band and optical range, 
but the two populations have similar hard X-ray fluxes and K magnitudes.
  \item The redshift distribution, built from a 50 count limit sample (119
AGN of which 73 with redshift), peaks at $z \sim$ 0.8. As previously noted 
by \cite*{JB-E1:ros01}, there is a strong excess of low $z$ AGN and a 
deficiency of sources at $z >$ 2 compared to population synthesis models 
for the X-ray background. The obscured/type 2 AGN cluster at 
$z \sim$ 0.8.
\end{itemize}
A similar analysis will be applied to the XMM-Newton deep survey of the CDFS 
(observations in progress) to characterize and classify the different AGN 
populations, and in particular to the objects optically too faint to get 
spectroscopic redshifts.

\begin{acknowledgements}

The results described here were made possible by the dedication of the 
XMM-Newton team in quickly producing well-calibrated Performance 
Verification data.  
G. Hasinger and I. Lehmann acknowledge support from the DLR grant 50 OR 9908.

\end{acknowledgements}


\begin{thebibliography}{}

\bibitem[\protect\astroncite{Barger et~al.}{2001}]{JB-E1:bar01}
Barger, A.J., Cowie, L.L., Mushotzky, R.F., \& Richards, E.A. 2001, AJ, 
121, 662

\bibitem[\protect\astroncite{Bruzual \& Charlot}{1993}]{JB-E1:bc93}
Bruzual, A.G., \& Charlot, S. 1993, ApJ, 405, 538

\bibitem[\protect\astroncite{Coleman et~al.}{1980}]{JB-E1:col80}
Coleman, G.D., Wu, C.C., \& Weedman, D.W. 1980, ApJS, 43, 393

\bibitem[\protect\astroncite{Giacconi et~al.}{2001}]{JB-E1:gia01}
Giacconi, R., Rosati, P., Tozzi, P.,  et al.\ 2001, ApJ, 551, 624

\bibitem[\protect\astroncite{Gilli et~al.}{2001}]{JB-E1:gil01}
Gilli, R., Salvati, M., \& Hasinger, G. 2001,  A\&A, 366, 407

\bibitem[\protect\astroncite{Granato et~al.}{1997}]{JB-E1:gra97}
Granato, G.L., Danese, L., \& Franceschini, A. 1997, ApJ, 486, 147

\bibitem[\protect\astroncite{Hasinger et~al.}{2001}]{JB-E1:has01}
Hasinger, G., Altieri, B., Arnaud, M., et al.\ 2001, A\&A, 365, L45

\bibitem[\protect\astroncite{Hasinger et~al.}{1998}]{JB-E1:has98}
Hasinger, G., Burg, R., Giacconi, R.,  et al.\  1998, A\&A, 329, 482 

\bibitem[\protect\astroncite{Hasinger et~al.}{1999}]{JB-E1:has99}
Hasinger, G., Lehmann, I., Giacconi, R.,  et al.\  1999, Proceedings of the
Symposium "Highlights in X-ray Astronomy" in honour of Joachim Truemper's 
65th birthday, eds. B. Aschenbach \& M.J. Freyberg, 1999, MPE Report 272, 
p. 199 [{\bf astro-ph/9901103}]

\bibitem[\protect\astroncite{Hornschemeier et~al.}{2001}]{JB-E1:hor01}
Hornschemeier, A.E., Brandt, W.N., Garmire, G.P., et al.\ 2001, ApJ, 554, 742

\bibitem[\protect\astroncite{Ishisaki et~al.}{2001}]{JB-E1:ish01}
Ishisaki, Y., Ueda, Y., Yamashita, A., et al.\ 2001, PASJ, 53, 445

\bibitem[\protect\astroncite{Koekemoer et~al.}{2001}]{JB-E1:koe01}
Koekemoer, A.M., Grogin, N.A., Schreier, E.J., et al.\ 2001, ApJ, in press 
[{\bf astro-ph/0110385}]

\bibitem[\protect\astroncite{Lehmann et~al.}{2001}]{JB-E1:leh01}
Lehmann, I., Hasinger, G., Schmidt, M.,  et al.\ 2001, A\&A, 371, 833 

\bibitem[\protect\astroncite{Lockman et~al.}{1986}]{JB-E1:loc86}
Lockman, F.J., Jahoda, K., \& McCammon, D. 1986, ApJ, 302, 432

\bibitem[\protect\astroncite{Mainieri et~al.}{2002}]{JB-E1:mai02}
Mainieri, V., Bergeron, J.,  Rosati, P., et al.\ 2002, in preparation

\bibitem[\protect\astroncite{Rosati et~al.}{2001}]{JB-E1:ros01}
 Rosati, P., Tozzi, P., Giacconi, R., et al.\ 2001, ApJ, in press 
[{\bf astro-ph/0110452}]

\bibitem[\protect\astroncite{Schmidt et~al.}{1998}]{JB-E1:sch98}
Schmidt, M., Hasinger, G., Gunn, J.,  et al.\  1998, A\&A, 329, 495

\bibitem[\protect\astroncite{Schreier et~al.}{2001}]{JB-E1:sch01}
Schreier, E.J., Koekemoer, A.M., Grogin, N.A.,  et al.\ 2001, ApJ, 560, 127

\bibitem[\protect\astroncite{Tozzi et~al.}{2001}]{JB-E1:toz01}
Tozzi, P., Rosati, P., Nonino, M., et al.\ 2001, ApJ, 562, 42

\bibitem[\protect\astroncite{Vanden Berk et~al.}{2001}]{JB-E1:van01}
Vanden Berk, D.E., Richards, G.T., Bauer, A., et al.\ 2001, AJ, 122, 549 

\bibitem[\protect\astroncite{Zamorani et~al.}{1981}]{JB-E1:zam81}
Zamorani, G., Henry, J.P., Maccacaro, T., et al.\ 1981, ApJ, 245, 357 

\end{thebibliography}
\end{document}